# Assessment of a high-resolution candidate detector for prostate time-of-flight positron emission tomography

Luigi Cosentino[1], Paolo Finocchiaro[1,a], Alfio Pappalardo[1], Franco Garibaldi[2]

[1]) *INFN, Laboratori Nazionali del Sud, 95125 Catania, Italy*

[2]) *INFN, Sezione di Roma1, Gruppo Collegato Sanità, 00161 Roma, Italy*

**Abstract.** We report on the measurements performed using a [22]Na source on a detector element for an MRI-compatible TOF-PET endorectal prostate probe, with Depth-Of-Interaction sensitivity. It is made from a LYSO scintillator crystal, wrapped with Lumirror, readout at both ends by means of Silicon Photomultipliers. With a detailed description of the data analysis procedure we show that our results point to a 400 ps coincidence resolving time and, at the same time, to a Depth-Of-Interaction resolution of 1 mm. These appealing features, along with the tiny 1.5 mm x 1.5 mm x 10 mm crystal size, are quite promising in view of the realization of a prototype probe.

[a] e-mail: FINOCCHIARO@LNS.INFN.IT

## I. INTRODUCTION

As it is well known prostate cancer is the second cause of cancer death in men, and a lot of efforts are currently aimed at early diagnosis and effective follow-up. In this respect a multimodal imaging approach could play a relevant role by merging anatomical and functional details from simultaneous Positron Emission Tomography (PET) and Magnetic Resonance Imaging (MRI) scans. The TOPEM project, funded by Istituto Nazionale di Fisica Nucleare in Italy (INFN), is going to develop an endorectal probe to perform Positron-Emission-Tomography (PET) in Time-Of-Flight (TOF) mode compatible with MRI[1]. It should be used in coincidence with an external detector array which could be planar or, better, a half PET ring[2,3]. The high quality timing performance of the probe will allow to considerably improve the Signal-to-Noise-Ratio (SNR) and the Noise Equivalent Counting Rate (NECR), thus boosting the image quality and the lesion detection capability[4,5,6,7]. Moreover, an increased sensitivity to the impact position of the gamma ray, known as Depth-Of-Interaction (DOI), improves the uniformity of the spatial resolution across the field of view, since the parallax error can be effectively compensated[8,9].

The probe aims at a coincidence resolving time of a few hundred picoseconds on collinear coincidences between the two gamma rays originated by the positron annihilation, by exploiting arrays of very compact scintillation detectors made from LYSO crystals. In this perspective, and due to its insensitivity to magnetic fields, the Silicon Photomultiplier (SiPM) represents an almost mandatory photosensor choice[10,11,12].

Even though TOF and DOI precision are somewhat mutually exclusive[13,14,15,16,17,18,19], in a recent paper we have shown that a suitable configuration of the detector element, along with specific algorithms for raw data analysis, might allow to overcome this limitation achieving an interesting tradeoff between detector element size, energy, time and DOI resolution[20]. Unfortunately, due to the lack of a proper positron source, those tests were performed with a collimated $^{137}$Cs gamma source and therefore the results could not be considered as a final proof.

In this paper we show, by means of tests performed with a $^{22}$Na positron source, that our extrapolation was basically correct and that it strongly supports the feasibility of our TOF-PET probe with the required features and performance.

## II. EXPERIMENTAL SET UP

The prototype probe we plan to build within the framework of the TOPEM project will be made of 450 (15x30) scintillator detectors, each one consisting of 1.5 mm x 1.5 mm x 10 mm LYSO crystals, with an overall active area of about 25 mm x 50 mm. At the moment LYSO seems the most promising scintillator because of its high average atomic number Z, good light yield and rather short light decay time ($\tau \approx 40$ ns). Its detailed features can be found in refs. 21,22,23. Each scintillator will have two SiPMs coupled to its ends, in order to allow the determination of the DOI while collecting as much scintillation light as possible thus optimising the energy resolution (see Fig. 1). The detector element we tested in this work is a single crystal, laterally wrapped with Lumirror[24] tape and readout by means of two Hamamatsu S10931-050P silicon photomultipliers. We remark that these 3mm x 3mm photosensors feature 3600 elementary square cells of 50 μm side, therefore on each side the crystal was only coupled to roughly 900 out of the 3600 cells.

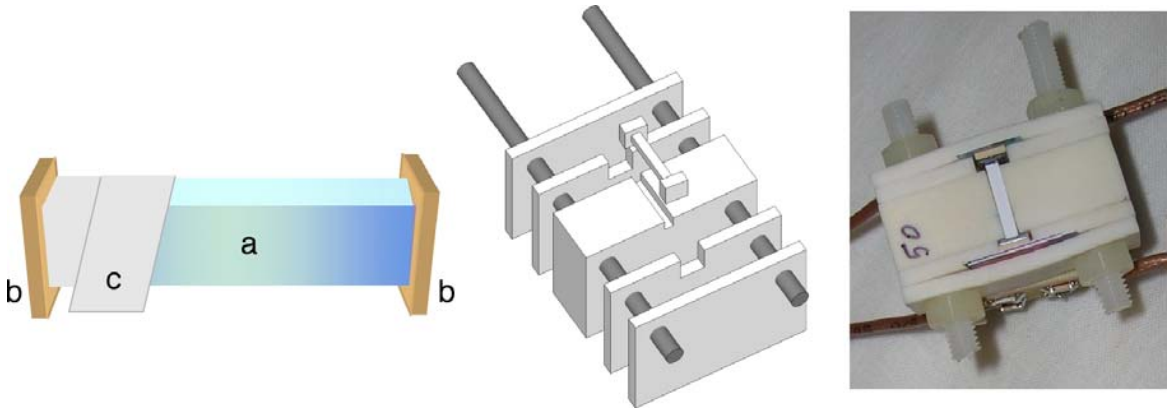

Fig. 1. (Color online) Lefthand side: sketch of the single detector element, made from a LYSO scintillator (a) readout by two SiPMs (b) and wrapped with Lumirror (c). Center and righthand side: the mechanical assembly of the tested prototype detector element, including the two SIPM voltage bias networks.

The advantage of using SiPMs, apart from their compactness, low bias voltage (30-70 V) and insensitivity to magnetic fields, is their very fast response (around 1 ns). Together with a fairly good photon detection efficiency, this promises an overall performance of timing and energy resolution comparable to (or even better than) high quality vacuum photomultipliers. The Photon Detection Efficiency (PDE) of these SiPMs around 400 nm is about 20%, as previously measured by us[25]. With calculations corroborated by suitable measurements we proved that, under these experimental conditions, the limited number of available microcells with the consequent non-linearity due to the multiple firing effect does not play a relevant role in the determination of the observables of interest. However, this topic goes beyond the scope of this work, and will be the subject of a dedicated forthcoming paper.

The role of each detector element consists in detecting the 511 keV gamma rays emitted by the radiotracer accumulated inside the tissue under examination, in coincidence with one of the external detectors. The probe must be able to allow a precision reconstruction of the position of the emitting source, as well as selecting the region of interest to be analyzed by means of the time of flight, thus improving the signal to background ratio[1,5,6].

In this work we concentrate on proving that our candidate detector element can attain satisfactory performance in terms of energy, time and DOI resolution, in view of the construction of a full size prototype probe. We remark that DOI reconstruction is mandatory in order to correct the data for the parallax effect caused both by an extended source and a planar detector array[9]. Several approaches have been pursued by various authors in order to optimize the DOI resolution, none of them so far definitively convincing, either for their cost, complexity and mechanical size, or for the overall performance[9,14].

Our approach consists in reading out the scintillation light from the two scintillator ends by means of SiPM photodetectors, and deducing the DOI by comparing the corresponding amounts of measured charge. Obviously, in order to be sensitive enough, this approach needs to differentiate the amount of transmitted light according to the longitudinal position of production (i.e. the gamma ray impact coordinate). This means that we need to be sensitive to the light attenuation along the crystal, thus implying that the crystal side-faces treatment and the reflector material must be chosen accordingly. This way the longer the path, the larger the average number of reflections needed, the smaller the amount of light reaching the photosensor. Unfortunately a considerable attenuation, even though benefitting the DOI sensitivity, spoils the attainable time resolution of a scintillator, as it depends on the photon collection statistics. In other words, one would like to reduce the fraction of transmitted photons per unit length in order to improve the DOI resolution and at the same time to keep it as large as possible in order to improve time resolution. We have found a reasonable tradeoff by adopting a scintillator with polished surfaces and a proper reflector (Lumirror) which, along with a suitable electronics and data analysis method, has allowed us to obtain performance that might represent a significant achievement in the technological development of compact TOF-PET devices. In order to perform the tests under PET realistic conditions we made use of a $^{22}$Na positron source that, after the positron annihilation, gives rise to two back-to-back 511 keV gamma rays. As we will see in the following chapters, we made our tests in three different geometrical configurations:

> Detection of single gamma rays, selected by means of lead collimators.

Coincidence of two gamma rays, detected in a LYSO+photomultiplier and our detector element placed horizontally with respect to the source (the gamma rays impinging perpendicular to the LYSO axis).

Coincidence of two gamma rays, detected in a LYSO+photomultiplier and our detector element placed vertically with respect to the source (the gamma rays impinging parallel to the LYSO axis).

### III. TEST WITH COLLIMATOR

In order to select the impact position of the 511 keV gamma rays coming from the $^{22}$Na source, we fabricated three lead collimators, 3 cm thick and 1 mm diameter. By means of these collimators we irradiated the scintillator in five different longitudinal positions, namely at 0 mm, ±2 mm, ±3.5 mm, referred to the midpoint, and measured the dependence of time and DOI on the position. A sketch of the experimental setup of source, collimator and detector is shown in Fig. 2, along with a picture of the three collimators employed.

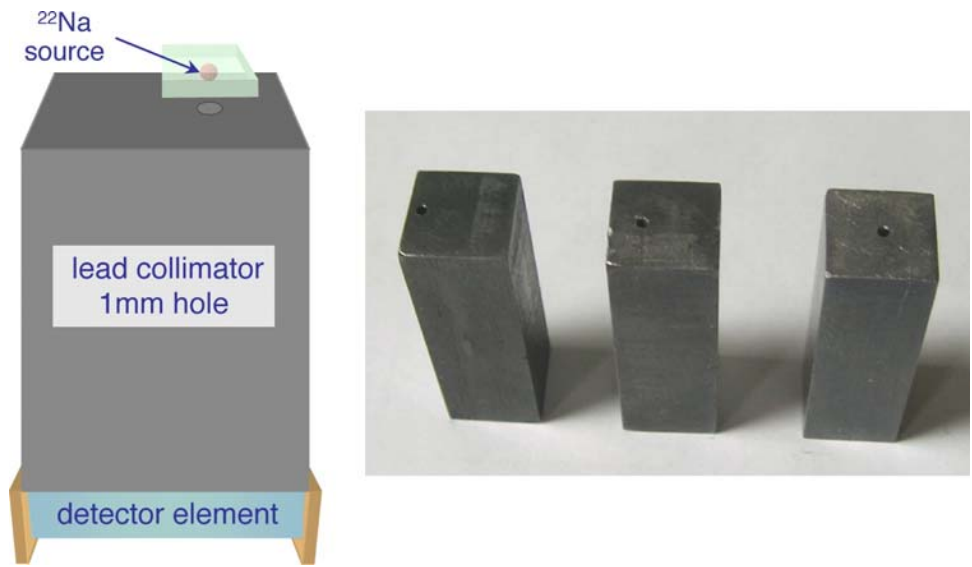

Fig. 2. (Color online) Lefthand side: sketch of the experimental setup of source, collimator and detector. Righthand side: the three lead collimators used for the 5 measurement positions.

The front end electronics included two fast voltage amplifiers whose outputs were split in two. A copy of these signals was used for charge-to-digital conversion (QDC, for energy and DOI measurements), whereas the other copy was fed into two leading edge discriminators and then used as start-stop signals for a time-to-digital converter (TDC).

In order to perform the measurements we made use of a home-made data acquisition system (DAQ) capable of digitizing and recording simultaneously the two charge signals produced by the SiPM sensors and the time interval between them. The DAQ trigger was set so that the three parameters ($Q_{left}$, $Q_{right}$, T) were only acquired whenever there was a left-right coincidence within a predefined time window that also created an amount of scintillation light beyond a minimum required threshold, thus enhancing the real gamma events while suppressing uncorrelated noise and spurious signals. A simplified logical scheme of the front-end and DAQ system is depicted in Fig. 3.

The total deposited energy, proportional to the total scintillation light produced, is:

$$E = k\sqrt{Q_L \cdot Q_R} \tag{1}$$

where $Q_L$ and $Q_R$ represent the charge values digitized from the left and right SIPM, and k is a calibration constant. For a more detailed explanation of (1) see ref.26.

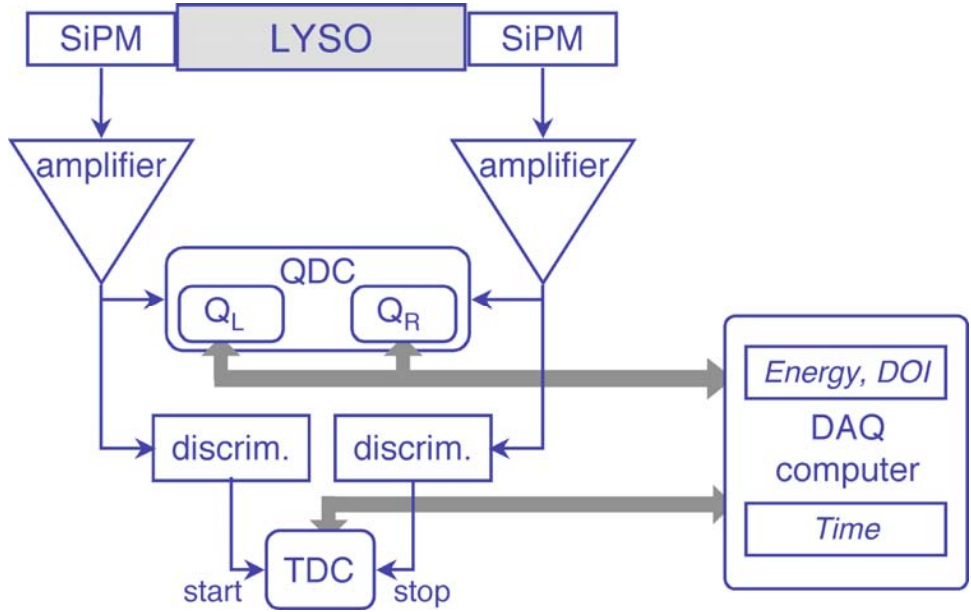

Fig. 3. (Color online) Simplified block scheme of the data acquisition system setup used for the test with collimators.

The left-right time difference was calibrated by correlating the values measured by the TDC when feeding its start and stop with the same pulser signal, being the stop delayed by precisely known time intervals by means of precision cable delay units. All of the measurements were performed with a time calibration of 12.4 ps/channel. Several preliminary measurements were done to choose the bias voltage for the SiPM, i.e. the one which minimizes the timing resolution, and the result was exactly the value recommended by the manufacturer.

For each of the five irradiation positions we built the ($Q_L$ vs $Q_R$) scatter plot, as shown in Fig. 4, where the full energy peaks are clearly visible and move along a well defined geometrical locus defined by E=$E_{peak}$ in (1).

By using (1) we built for each irradiation position the non-calibrated energy histogram, as shown in Fig. 5 along with their sum and a fit. The peak at 511 keV is clearly visible despite the considerable background due to scattering inside the lead collimator, and has an FWHM resolution (in channels) around 13%. In the following, whenever we select full-energy events it will mean choosing events whose energy value from (1) falls within an FWHM window around this peak. In the same figure one observes a systematic slight displacement of the full-energy peak while moving from the sides to the center of the crystal, this implying a slight deviation from the ideal behavior expressed by (1). However, as high energy resolution is not needed, this does not spoil the overall performance of the detector.

It is worth to remark that in the spectrum of Fig. 5 one can also see the additional gamma peak at 1274 keV, along with its Compton spectrum, that disappears when the measurements are performed in coincidence as will be shown in the following chapter.

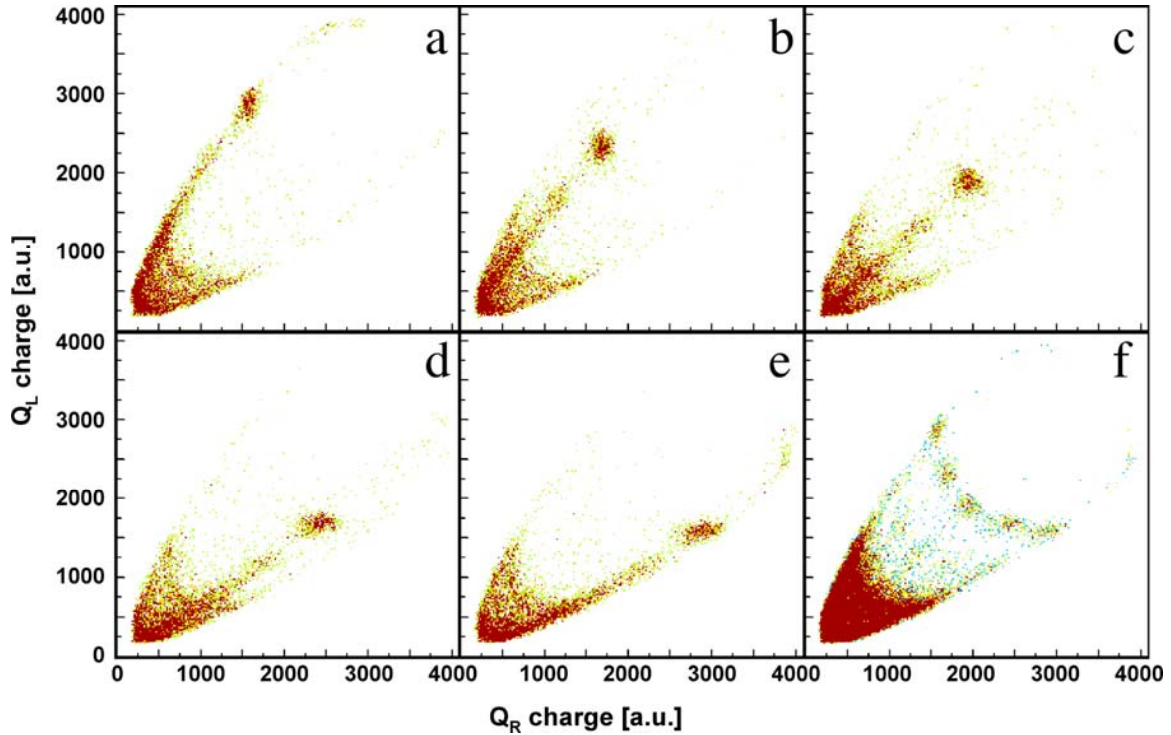

Fig. 4. (Color online) (a, b, c, d, e) Scatter plots ($Q_L$ vs $Q_R$) for the five irradiation positions. The 511 keV full energy peak is clearly visible, and it moves along a locus defined by E= $E_{peak}$ in (1). (f) The superposition of the previous five scatter plots.

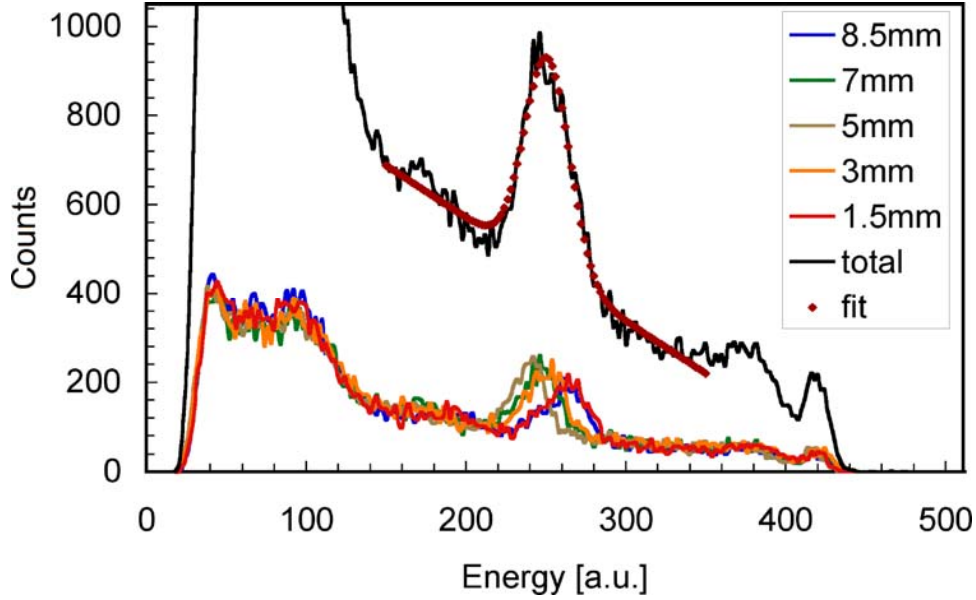

Fig. 5. (Color online) Energy spectrum for the five irradiation positions, obtained by means of (1), and their overall sum. The FWHM resolution around the 511 keV peak (computed in channels) is around 13%.

The calculation of DOI was done by means of the following formula:

$$DOI_L = M \frac{Q_L}{Q_L + Q_R} \qquad (2)$$

where M is a suitable calibration constant to be determined experimentally by correlating the measured full-energy data with the known irradiation positions via the collimators. The same considerations apply in exactly a symmetrical fashion to the case of $DOI_R$, as the two quantities are strongly bound to each other. From now on we will only use $DOI_L$ for our calculations and assumptions, and will denote it simply DOI. Notice that (2) provides DOI in the range 0-10 mm, therefore in this coordinate system, that will be used henceforth, the five positions defined by the collimators become respectively 1.5, 3, 5, 7, 8.5 mm.

However, in order for (2) to hold, $Q_L$ and $Q_R$ must be homogeneous, thus one could decide to equalize the response of the SiPMs as best one can, so that the charge read by the two QDC channels when the source in on the middle position is almost the same. For instance one could bias the SiPMs at the same value and use identical amplifier channels, but small differences in the total gain between the two energy channels will be present anyhow. Even a tiny left-right difference in the optical coupling quality will introduce differences in the output amplitudes, and therefore we decided that the best solution was to equalize $Q_L$ and $Q_R$ in software. Thus event by event we multiplied $Q_R$ by a suitable normalization constant, calculated in order to make the scatter plot of Fig. 4 symmetric.

In Fig. 6 we show the distribution of the DOI values as a function of the raw position and after calibration in millimeters done by means of Fig. 7.

Once calibrated the DOI, we were interested in knowing the precision in its measurement. The width of each gaussian in Fig. 6, that is the statistical error to be attributed to each DOI measurement performed on the occurrence of one full-energy gamma detection, was reported in Fig. 8 together with the theoretical expectation by using the accurate formula of ref.27. As can be seen the FWHM in each position is around 1mm or smaller, in nice agreement with the expected values from theory and from the 1mm collimator diameter.

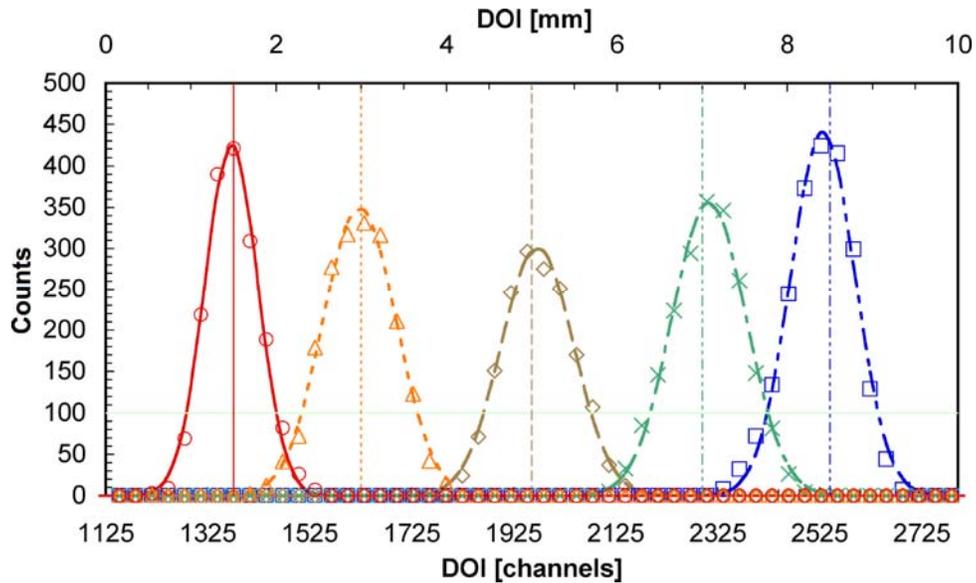

Fig. 6. (Color online) Distribution of the DOI values when separately irradiating the crystal onto each of the five predetermined positions. The symbols represent the data points, the bell-shaped curves are gaussian fits, the vertical lines are the nominal collimator positions.

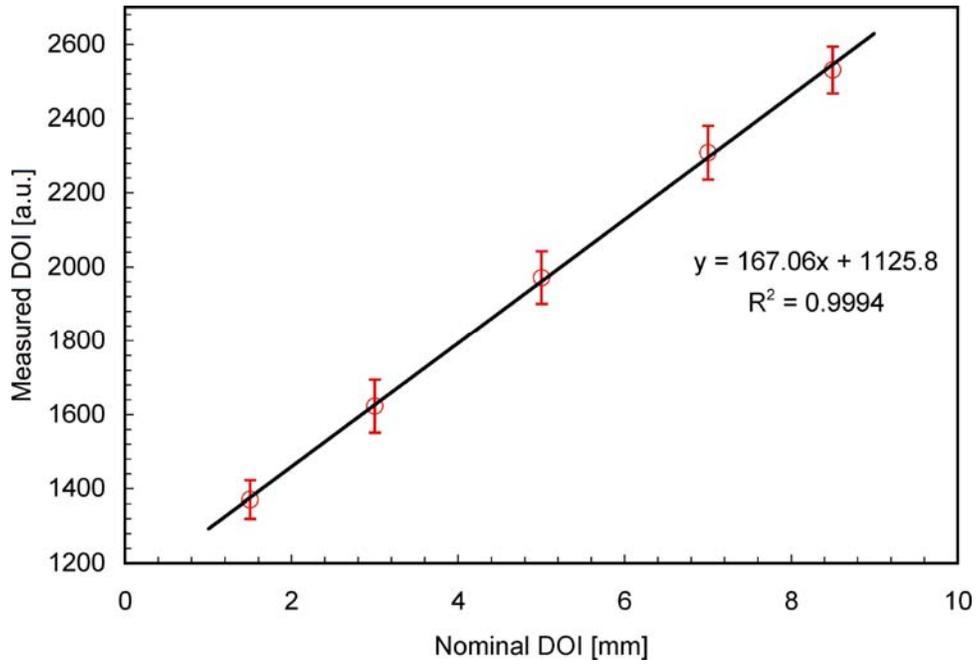

Fig. 7. (Color online) DOI computed using (2) as a function of the nominal collimator position.

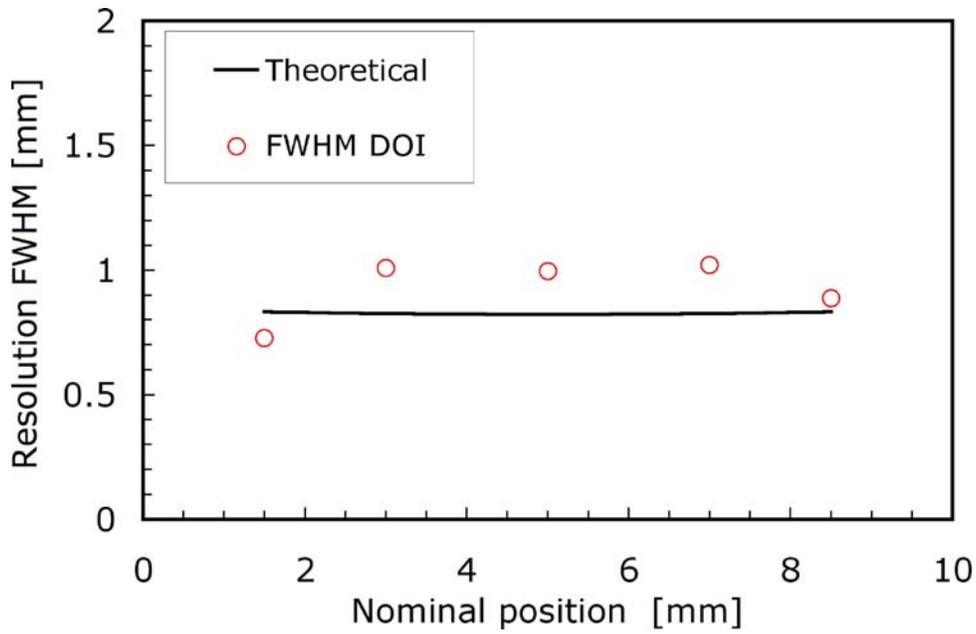

Fig. 8. (Color online) DOI resolution as a function of the nominal collimator position. The theoretical value, below 1 mm, was derived using the accurate formula in ref. 27.

In order to study the detector performance in terms of timing we constructed the five histograms representing the distribution of the left-right time difference as detected by the two SiPMs. These plots, built under the condition $\left|E - E_{peak}\right| \leq \frac{1}{2}\Delta E_{FWHM}$ and shown in Fig. 9, are narrow gaussian-shaped distributions ($\Delta t_{FWHM} \approx 400$ ps) whose individual widths are plotted in Fig. 10. These curves are displaced with respect to each other, the reason being the different propagation time of the light signal inside the crystal itself, as a function of the impact position. The time shift with DOI position was linearly corrected event by event by making up for the different propagation times. The linear correction coefficients can be easily deduced by fitting the position of the centroids of Fig. 9 as a function of the nominal collimator position, as shown in Fig. 11 which also allows to deduce the light signal propagation speed inside the crystal. After applying the just mentioned correction we rebuilt the timing plot, also calculating the overall sum plot as shown in Fig. 12, with an FWHM resolution around 420 ps. The goodness of such a correction can also be seen on the time versus DOI scatter plot shown in Fig. 13: after the correction there is no longer any dependence of time on the impact position. We remark that by assuming identical behavior of the two SiPMs one can infer an individual contribution $\Delta t_R = \Delta t_L \approx 420/\sqrt{2} \approx 300\,ps$, in reasonable agreement with the *285 ps* expected value as extrapolated in ref.20.

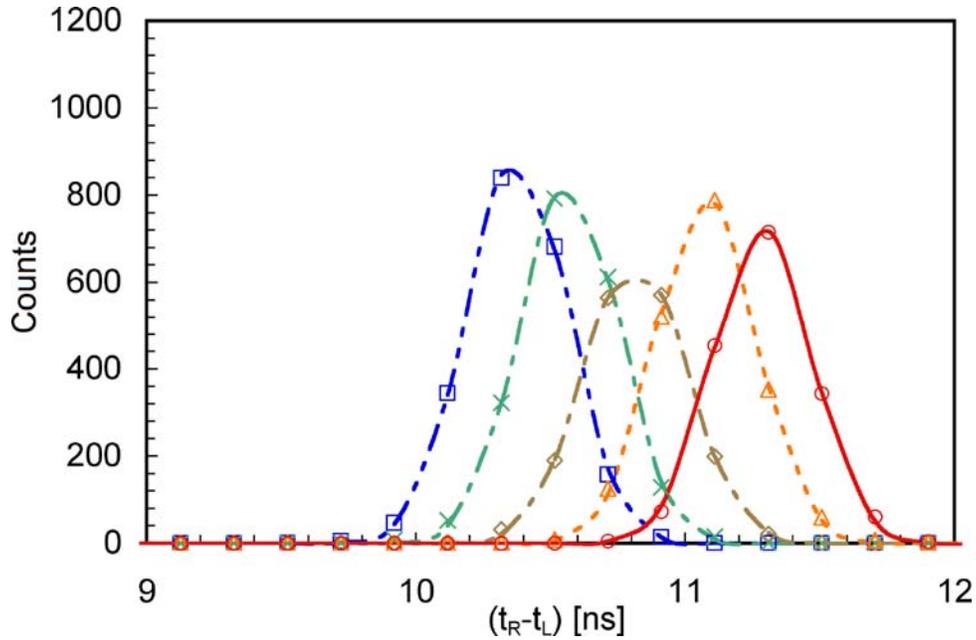

Fig. 9. (Color online) Time spectrum ($t_{right}$-$t_{left}$) for full-energy peak events originated in the five selected positions. The gaussian shaped distributions are narrow but displaced with respect to each other.

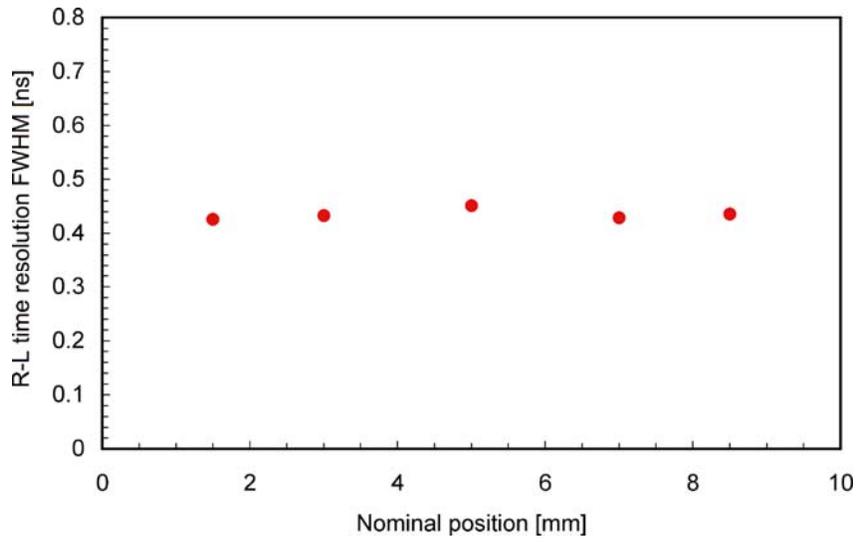

Fig. 10. (Color online) Measured FWHM time resolution (i.e. the widths of the five gaussians of Fig. 9) as a function of the nominal collimator position.

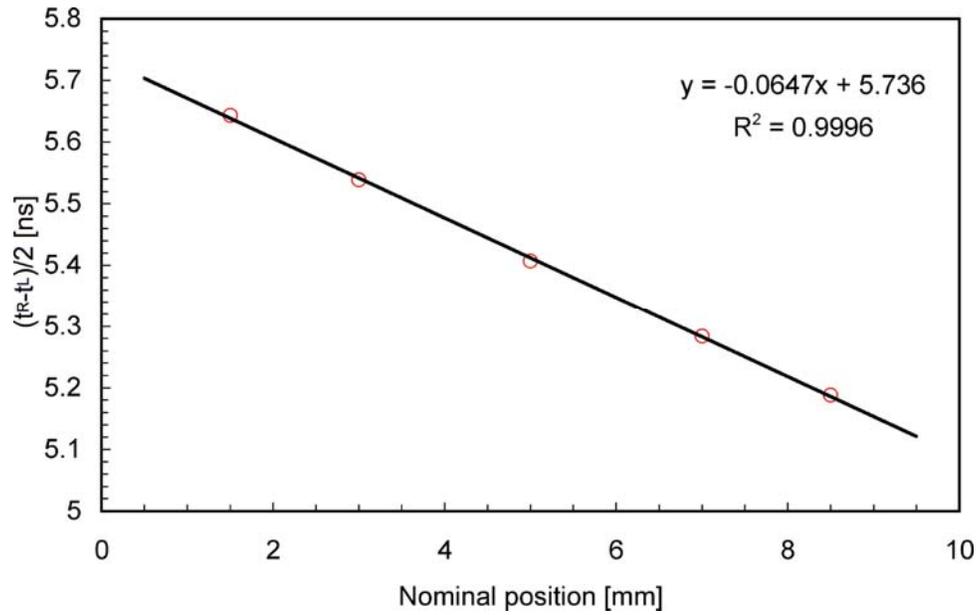

Fig. 11. (Color online) $(t_R-t_L)/2$ as a function of the nominal collimator position. From the slope of the linear fit we deduced the propagation speed of the light signal as *v=15.5 mm/ns*.

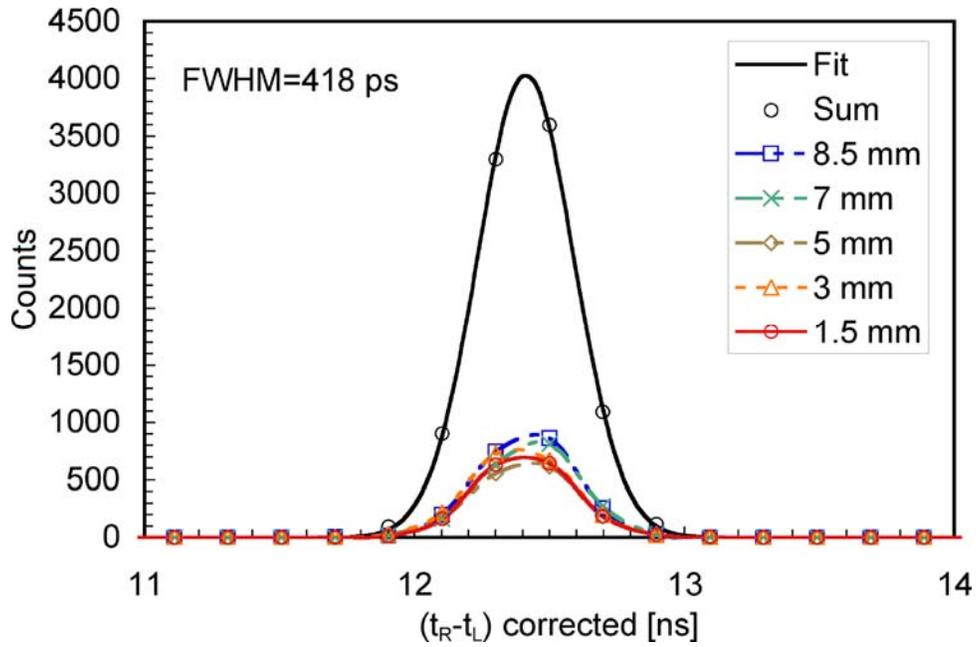

Fig. 12. (Color online) The same plots of Fig. 9 after correcting for the time shift due to the light-signal finite propagation speed inside the crystal. The overall sum plot is also shown, whose FWHM resolution is 418 ps.

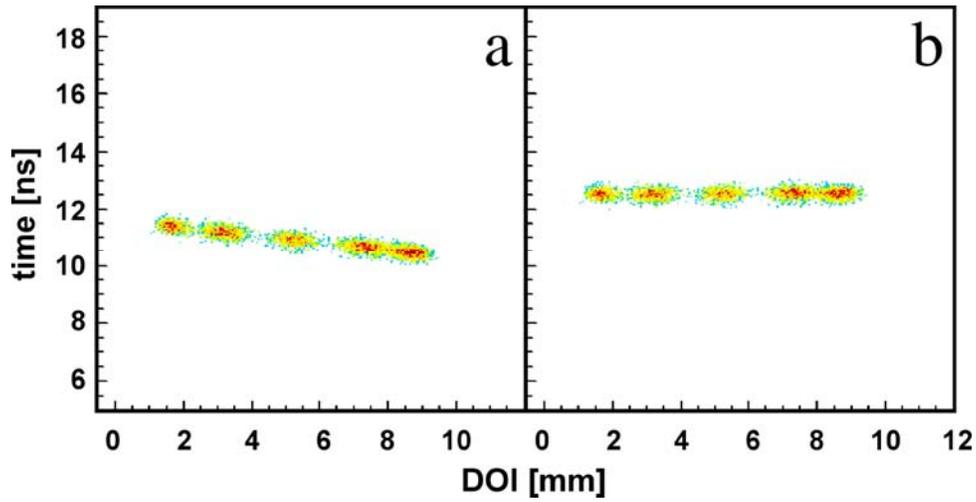

Fig. 13. (Color online) (a) Time versus DOI scatter plot for the five selected impact positions. The downshift of the time values as the impact position increases is clearly seen. In (b) the same plot after correcting for the light-signal propagation speed inside the crystal.

## IV. TEST WITH PMT AND HORIZONTAL DETECTOR

For this test we made use of an additional detector, namely a 3 mm x 3 mm x 3 mm LYSO coupled to a 9133B photomultiplier tube (PMT) produced by Electron Tubes, in order to select event based on the geometrical constraint of the two 511 keV gamma rays emitted simultaneously at 180° relative angle. Our detector element was placed horizontally with respect to the PMT face and to the $^{22}$Na source, as shown in Fig. 14. The distances between the source and the scintillators were respectively 9 cm and 1 cm, in order to select an interaction spot of about 1 mm size onto the detector element under test.

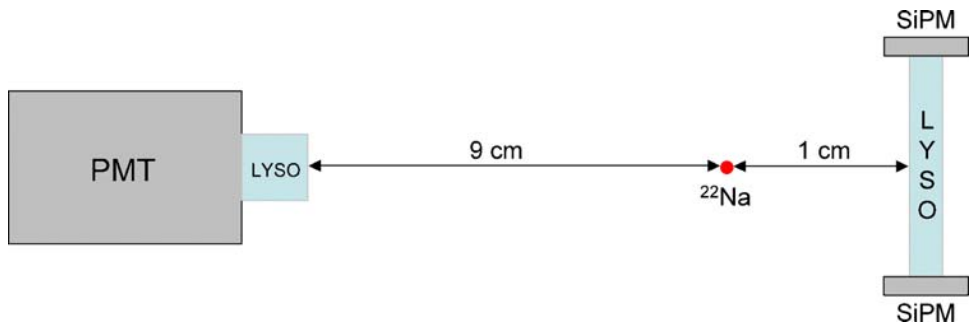

Fig. 14. (Color online) Geometrical arrangement of the measurement performed in coincidence between LYSO+PMT and our detector element in "horizontal" position with respect to the source (drawing not in scale).

The front end electronics was quite similar to the previous ones, the difference being that in this case the trigger condition was the simultaneous detection of light by the two SiPMs and the PMT. As for the data acquisition we made use of three Time-to-Digital Converters in order to record the time differences between each pair of photosensors. The simplified logical scheme of the front-end and DAQ system is shown in Fig. 15.

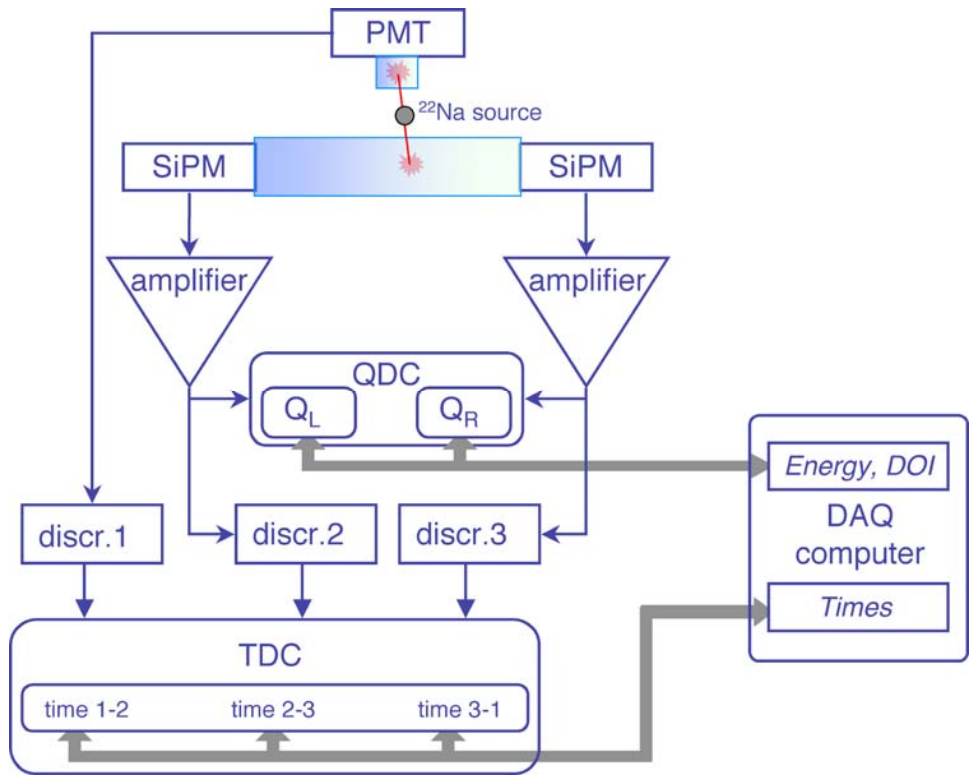

Fig. 15. (Color online) Simplified block scheme of the data acquisition system setup used for the coincidence measurements between the PMT and the LYSO.

We performed three sets of measurements by displacing the detector element along its axis, in order to have it irradiated in three different positions. As the effective setup was rather bulky, due to the small room available inside the dark box hosting the detectors and the amplifiers, the possible displacements were limited. However, the irradiation spots were the midpoint and two side positions displaced respectively by 2.5 mm and 3 mm. A visual check of this operating procedure is provided by the overall ($Q_L$ vs $Q_R$) scatter plot of Fig. 16, where one can immediately see the presence of three regions corresponding to the three irradiation positions. The full energy peaks are clearly visible, as well as a relevant reduction of the low-energy background due to the absence of scattering from the lead collimator and to the cleaning effect of the coincidence with the PMT.

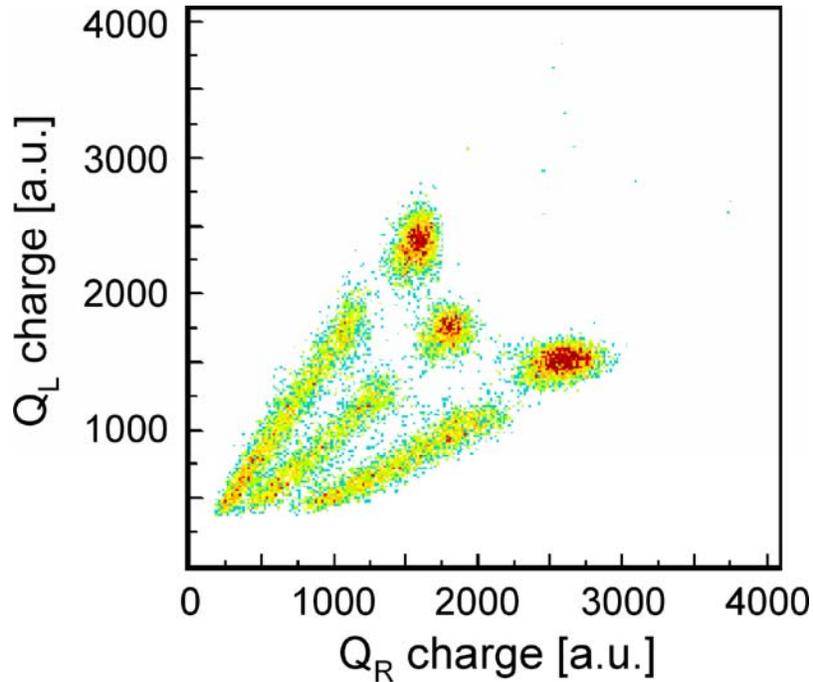

Fig. 16. (Color online) Overall scatter plot ($Q_L$ vs $Q_R$) for the three irradiation positions in the "horizontal" setup. The 511keV full energy peak is clearly visible, and it moves along a locus defined by E=constant in (1). The absence of the lead collimator and the coincidence with the PMT strongly suppressed the background counts.

This is also quite evident in Fig. 17, where one can also observe the disappearance of the 1274 keV peak and its tail, produced by the $^{22}$Na source and visible in Fig. 5, because of the coincidence trigger. Due to the same reasons an improvement is also visible in the DOI resolution, as can be seen in the bell-shaped distributions of Fig. 18 whose FWHM widths are reported in Fig. 19 along with the theoretical expectation again from ref. 27. We remark that the DOI calibration obtained in the configuration with the collimator holds also in this configuration, as expected. The DOI resolution improvement also reflects into a time resolution improvement, as can be seen in Fig. 20 where we reported the ($t_{right}$-$t_{left}$) distribution after applying the previously explained correction for the time shift due to the light-signal finite propagation speed inside the crystal. The FWHM width is now reduced from 418 ps to 404 ps.

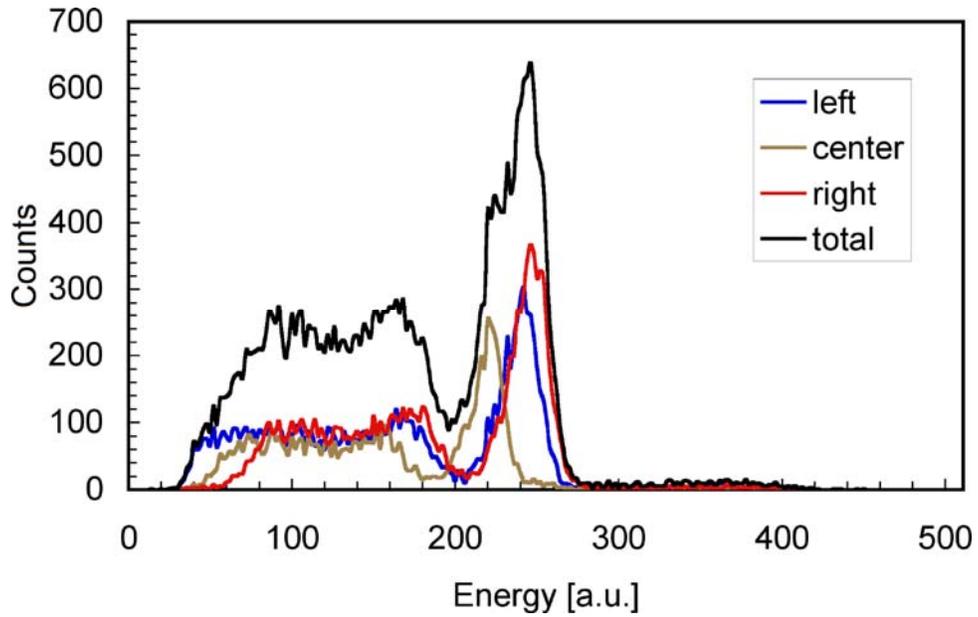

Fig. 17. (Color online) Energy spectrum for the three irradiation positions, obtained by means of (1), and their overall sum in the "horizontal" setup. The background due to scattering in the lead collimator has now disappeared, as well as the high-energy tail up to the peak at 1274 keV suppressed by the coincidence trigger.

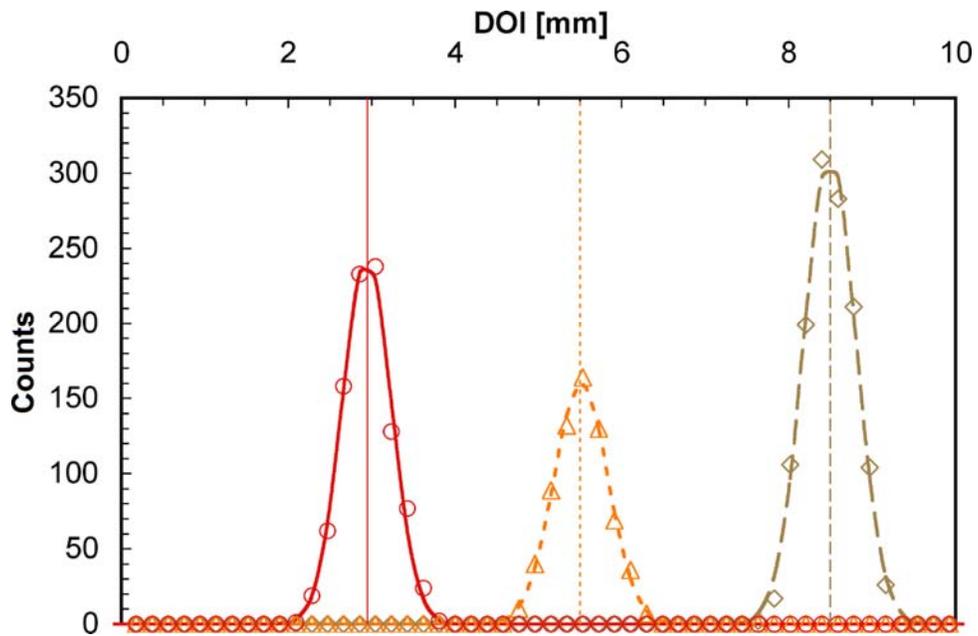

Fig. 18. (Color online) Distribution of the DOI values when separately irradiating the crystal onto the three selected positions ("horizontal" setup). The symbols represent the data points, the bell-shaped curves are gaussian fits, the vertical lines are the centroid positions.

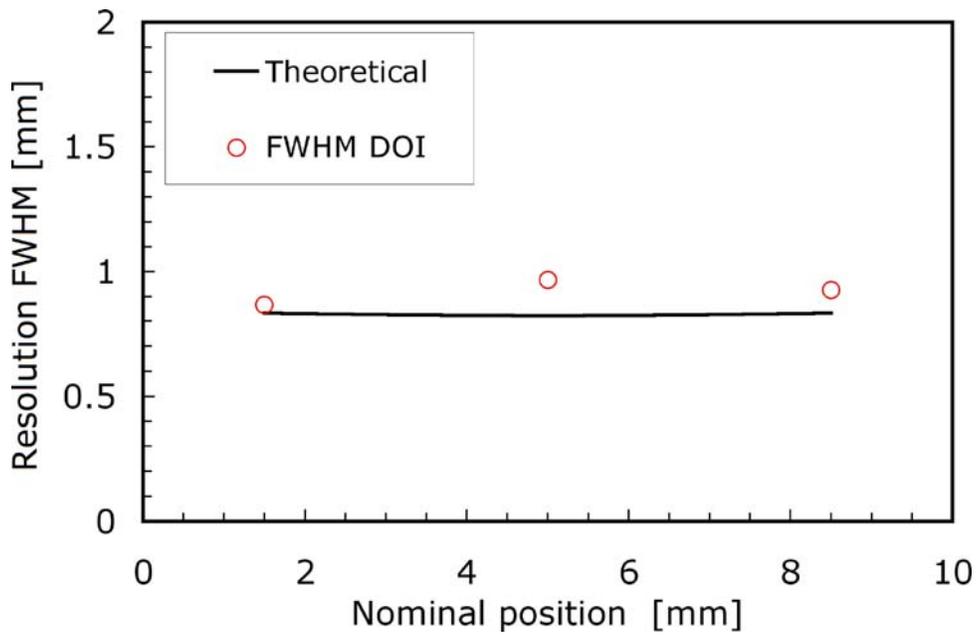

Fig. 19. (Color online) DOI resolution as a function of the nominal impact position ("horizontal" setup). The theoretical expectation was derived using the accurate formula in ref. 27.

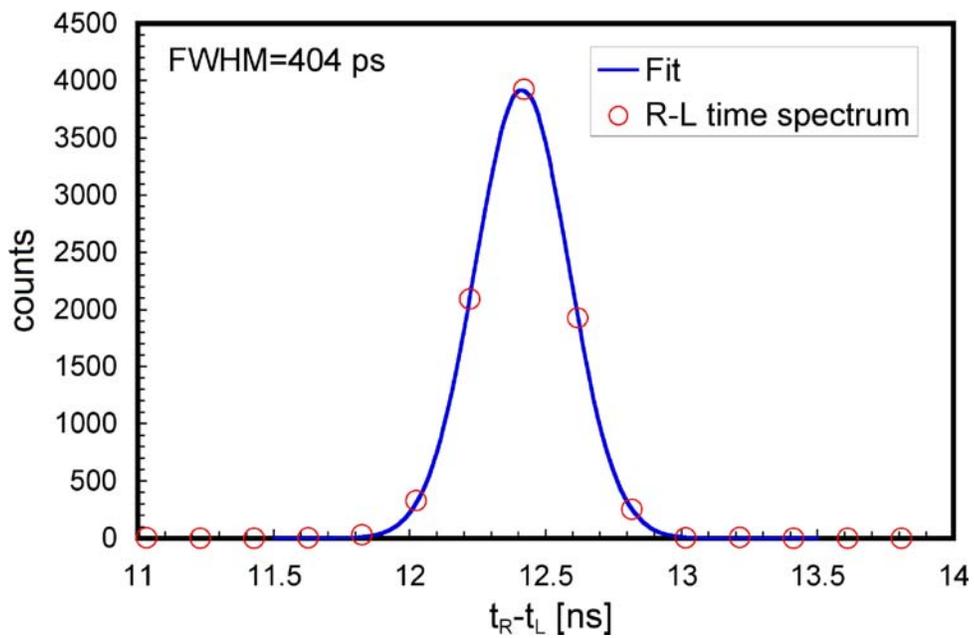

Fig. 20. (Color online) The ($t_{right}$-$t_{left}$) time distribution, after correcting for the time shift due to the light-signal finite propagation speed inside the crystal ("horizontal" setup).

## V. TEST WITH PMT AND VERTICAL DETECTOR

The setup for this test was identical to the previous one, the only difference being the orientation of the detector element in what we called "vertical" position. This is schematically sketched in Fig. 21 (the drawing is not in scale). The distances between the source and the scintillators were respectively 8 cm and 1 cm, and the detector element under test was irradiated across one of the two SiPMs, exactly in the fashion it would be in a realistic PET system.

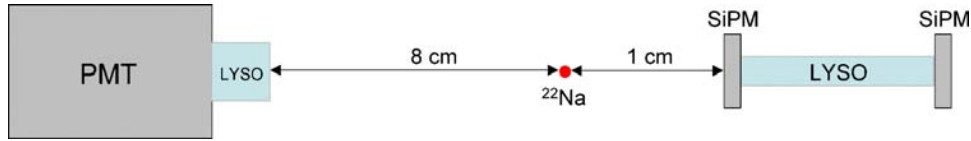

Fig. 21. (Color online) Geometrical arrangement of the measurement performed in coincidence between LYSO+PMT and our detector element in "vertical" position with respect to the source (drawing not in scale).

In such a configuration we no longer had the possibility of cross checking the measured DOI against nominal values, therefore we relied on the previous calibration that was valid for both the "collimator" and the "horizontal" configurations discussed above. In Fig. 22 we reported the ($Q_L$ vs $Q_R$) scatter plot, where one can now observe the full distribution due to the simultaneous irradiation of the whole detector element in the axial direction. Again, the full energy peaks are clearly visible in form of a hyperbolic locus defined by (1), as well as a relevant reduction of the low-energy background due to the absence of scattering from the lead collimator and to the cleaning effect of the coincidence with the PMT.

By making use again of (1) we built the energy spectrum reported in Fig. 23 which allows a clean selection of the full-energy 511 keV peak. By selecting events under such a peak, and applying the previously explained correction for the time shift due to the light-signal finite propagation speed inside the crystal, we built once again the ($t_{right}$-$t_{left}$) time distribution and plotted it in Fig. 24. The FWHM width in this case, consistently with the two previously tested configurations, is 427 ps.

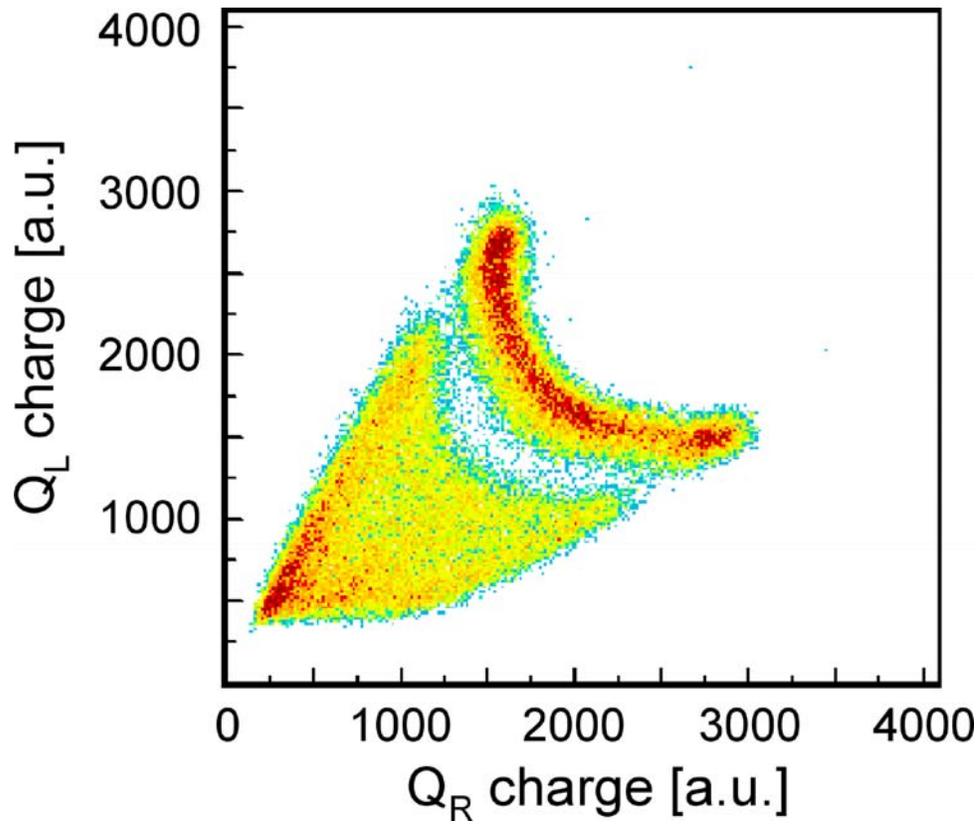

Fig. 22. (Color online) Overall scatter plot ($Q_L$ vs $Q_R$) for the "vertical" setup. The 511 keV full energy peak is clearly visible, and it totally covers a locus defined by E=constant in (1). Again, the absence of the lead collimator and the coincidence with the PMT strongly suppressed the background counts.

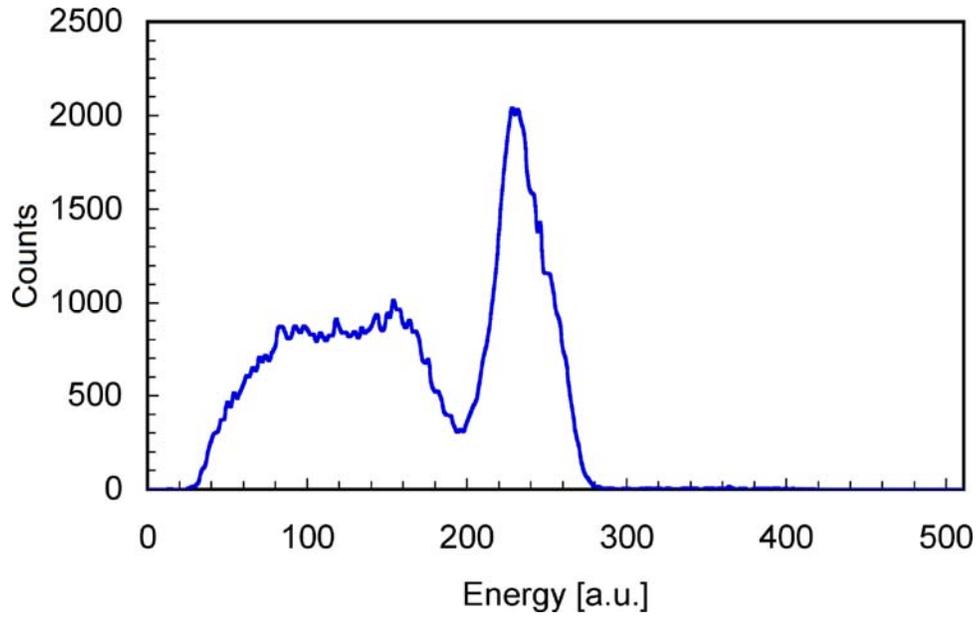

Fig. 23. (Color online) Energy spectrum for the "vertical" setup, obtained by means of (1). The background due to scattering in the lead collimator has disappeared, as well as the high-energy tail up to the peak at 1274 keV suppressed by the coincidence trigger.

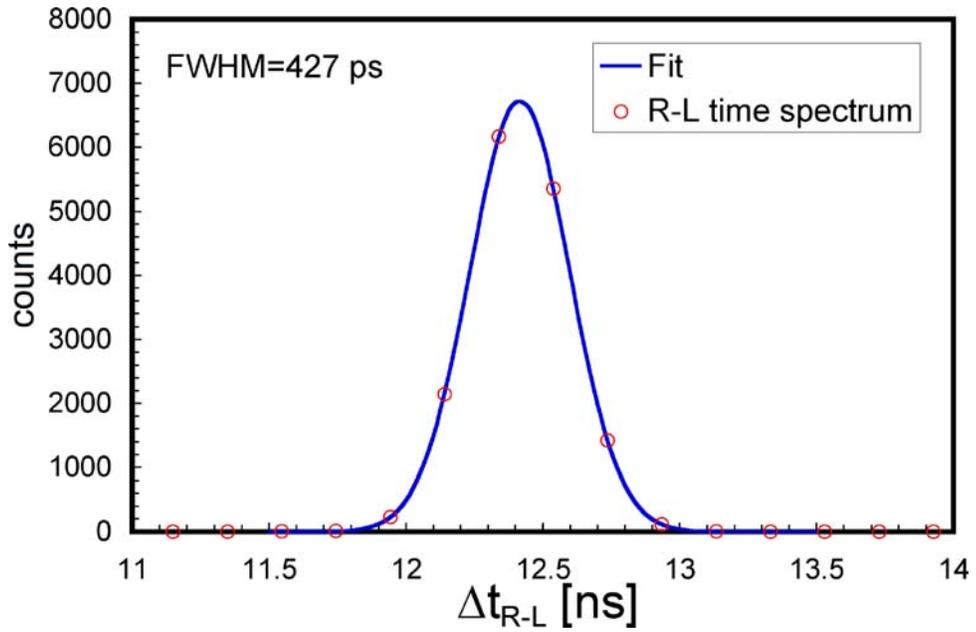

Fig. 24. (Color online) The ($t_{right}$-$t_{left}$) time distribution, after correcting for the time shift due to the light-signal finite propagation speed inside the crystal ("vertical" setup).

## VI. DISCUSSION

The tests performed under the three different configurations make us confident that the performance of our detector element in terms of energy, DOI and time are rather reproducible and do not depend appreciably on the detector orientation with respect to the impinging gamma rays. Moreover, even the first test, performed without a coincidence trigger and with a huge scattering background due to the presence of a lead collimator, gives basically the same performance. These results are also consistent with our previous findings on another detector, after proper rescaling due to different SiPMs with different number and size of cells and different PDE, different photon statistics due to the use of a $^{137}$Cs source[20].

By making use of the two additional times recorded in the coincidence measurements (horizontal and vertical setups) we now want to assess the detector performance in terms of Coincidence Resolving Time (CRT), that is a key parameter for the TOF-PET. With the naming scheme illustrated in Fig. 25, where $v$ is the light-signal propagation speed in the crystal, we have that:

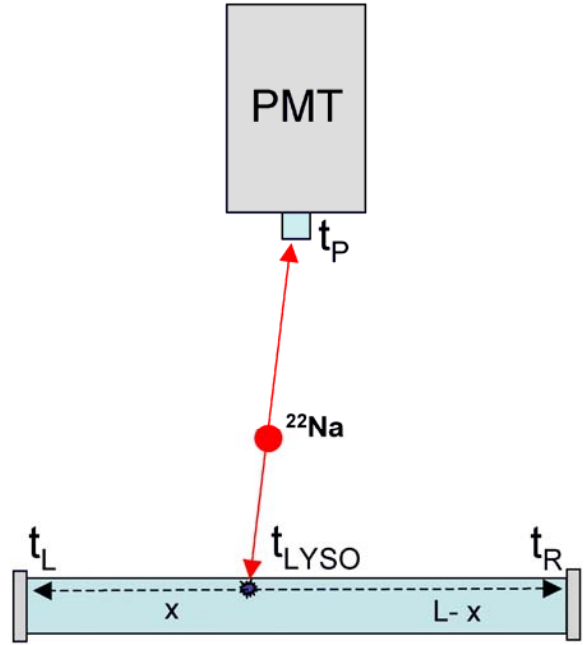

Fig. 25. (Color online) Naming scheme for the evaluation of the Coincidence Resolving Time when operating in coincidence.

$$t_{LP} = t_L - t_P = t_{LYSO} + x/v - t_P \qquad (3)$$

$$t_{RP} = t_R - t_P = t_{LYSO} + (L-x)/v - t_P \qquad (4)$$

By summing (3) and (4) one obtains:

$$t_{LYSO-P} = t_{LYSO} - t_P = \frac{t_{LP} + t_{RP}}{2} + \frac{L}{2v} \qquad (5)$$

We built event by event the $TOF = \dfrac{t_{LP} + t_{RP}}{2}$ variable both for the "horizontal" and the "vertical" configuration; the corresponding histograms, shown in Fig. 26 and Fig. 27, have FWHM widths respectively of 386 ps and 403 ps. In light of these results we can safely state that the coincidence resolving time of the system constituted by the PMT and the detector element under study is

$$\Delta_{TOF} = CRT \approx 400\, ps \qquad (6)$$

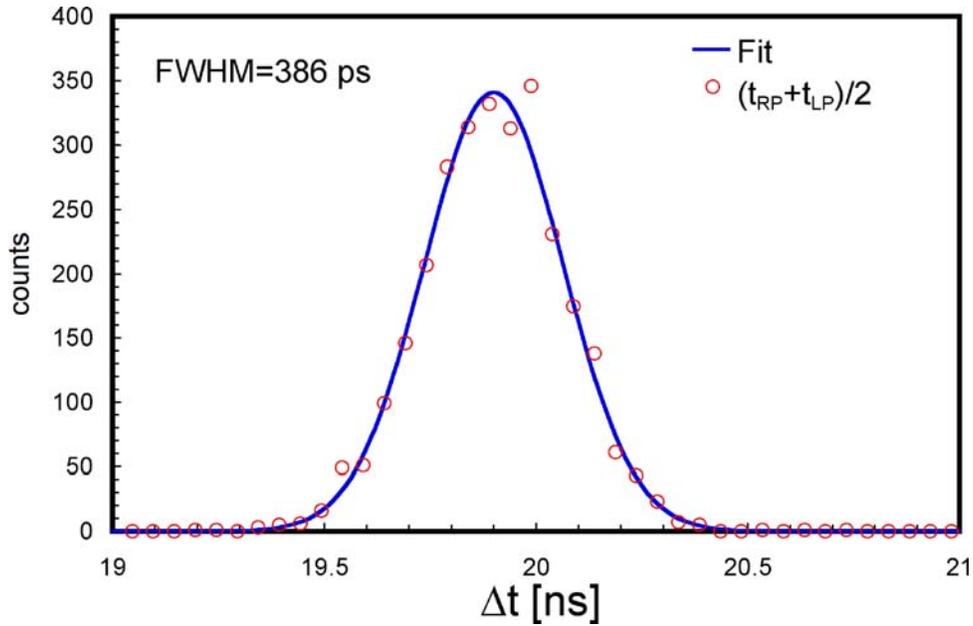

Fig. 26. (color online) Distribution of the $(t_{RP}+t_{LP})/2$ variable built event by event in the "horizontal" setup. Its FWHM width of 386 ps is the coincidence resolving time.

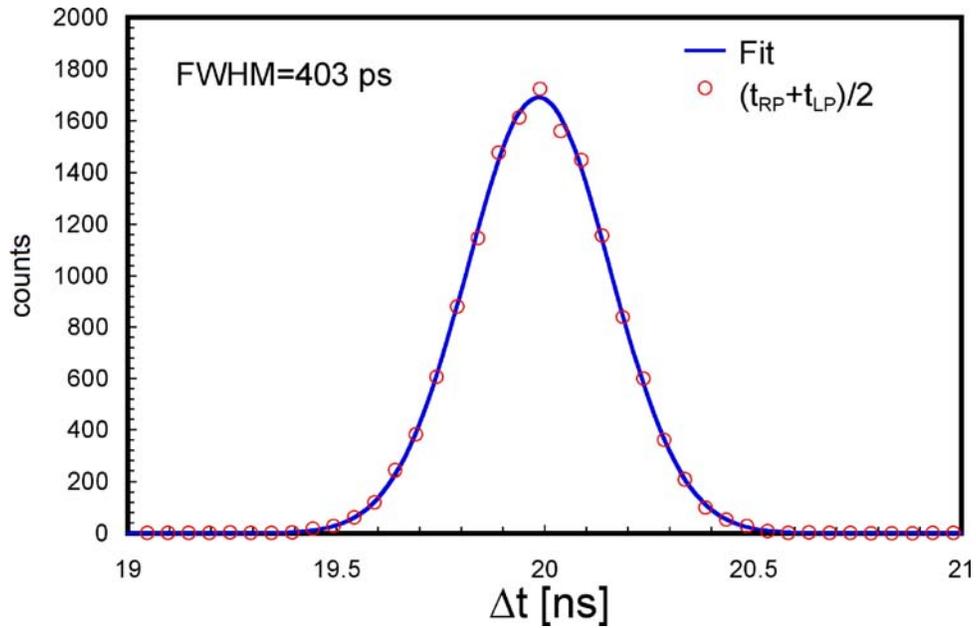

Fig. 27. (Color online) Distribution of the $(t_{RP}+t_{LP})/2$ variable built event by event in the "vertical" setup. Its FWHM width of 403 ps is the coincidence resolving time.

It is remarkable that even in case of no selection on the full-energy peak, i.e. by using all the recorded events, the TOF was basically independent of the DOI, as expected and as can be seen by comparing Fig. 28a and Fig. 28b. Obviously, if one does not make any energy selection (Fig. 28a) the TOF distribution widens due to a worse photon statistics.

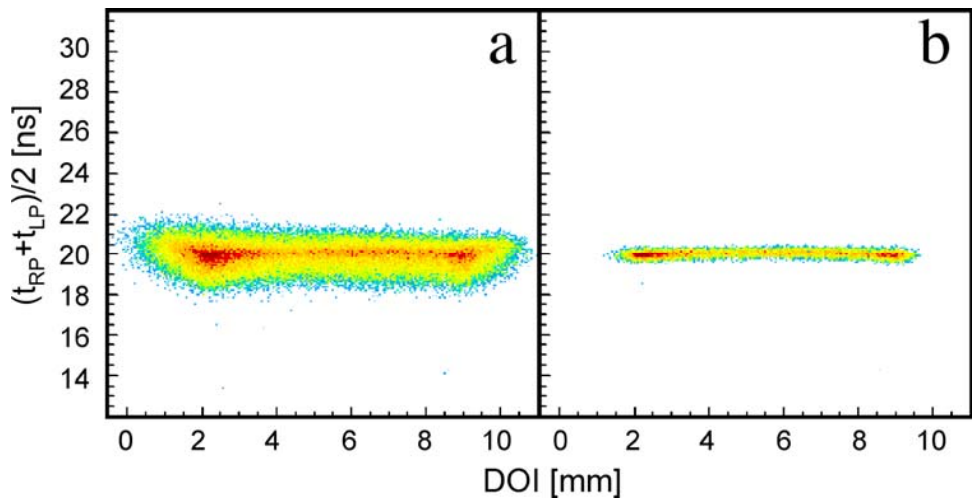

Fig. 28. (Color online) Scatter plot of time versus DOI. (a) without selection on the full-energy 511 keV peak; (b) with selection on the full-energy 511 keV peak. The TOF is independent of DOI as expected in both cases.

If *r* is the distance along the LOR from the midpoint between the two detectors to where the positron annihilation took place, and *c* is the speed of light, one gets:

$$r = c \frac{t_{LYSO-P}}{2} = c \frac{t_{LP} + t_{RP}}{4} + \frac{L}{4v} = c \frac{TOF}{2} + \frac{L}{4v} \tag{7}$$

and therefore the spatial resolution

$$\Delta r = c \cdot \frac{CRT}{2} \approx c \cdot 200\, ps \approx 6 cm \quad FWHM \tag{8}$$

which allows to estimate a 6 cm segment along the LOR as the origin of the positron annihilation into two gamma rays. With the probe we are going to build, at the moment of the PET image reconstruction one will then set a position window Δ*r* along the LOR, thus restricting the number of effective coincidences and consequently improving the Signal-to-Noise Ratio and the Noise Equivalent Counting Rate.

## VII. CONCLUSION AND PERSPECTIVES

With our measurements we have proved that the goal of the TOPEM project, in terms of the single detector element performance, can be achieved. The tests, performed under three different geometries, with and without a lead collimator, have shown stable results. In particular for the first time, as far as we know, a 400 ps coincidence resolving time was obtained simultaneosly with a 1 mm precision in depth of interaction. As we know that the timing performance of the PMT/LYSO detector was worse than the 2-SiPM/LYSO, we are confident that by replacing it with another 2-SiPM/LYSO the coincidence resolving time will further improve. We foresee to assemble a dedicated setup for these additional measurements as next step for the near future.

In our opinion these results, together with the 1.5 mm pitch of the detector element, represent an important step towards the construction of a high-performance TOF-PET miniaturized probe for the prostate cancer diagnosis. Moreover, such a kind of detector could also open new perspectives for applications to other organs, like breast and brain.

## VIII. ACKNOWLEDGMENTS

We are indebted to several people of the INFN-LNS staff, namely: C. Calì, P. Litrico, S. Marino, G. Passaro for their support with electronics; F. Ferrera for his help with the data acquisition system; B. Trovato, S. Di Modica, M.

Tringale, G. Vasta for all the aspects related to micromechanical machining. It is worth to mention G. Di Venti who did his thesis work on the detector described in this paper.